\documentclass[aps,onecolumn,showpacs,showkeys]{revtex4}
\usepackage[T1]{fontenc}
\usepackage{array}
\usepackage{yfonts}
\usepackage{mathrsfs}
\usepackage{graphicx}
\usepackage{latexsym}
\usepackage{amsmath}
\usepackage{amssymb}
\usepackage{hyperref}
\usepackage{natbib}
\usepackage[english]{babel}
\def \be{\begin{equation}}
\def \ee{\end{equation}}
\def \bea{\begin{eqnarray}}
\def \eea{\end{eqnarray}}
\begin{document}
\title{Physical state condition in Quantum General Relativity as a consequence of BRST symmetry}
\date{\today}
\author{Michele Castellana$^{\,1}$}
\author{Giovanni Montani$^{\,2,\,3,\,4}$}
\affiliation{$^1$Dipartimento di Fisica, Universit\`a  di Roma, ``Sapienza'', Piazzale A. Moro, 5, 00185 Rome, Italy.}
\affiliation{$^2$ENEA C.R. Frascati (Dipartimento F.P.N.), Via Enrico Fermi, 45, 00044 Frascati (Rome), Italy.}
\affiliation{$^3$ICRANet C. C. Pescara, Piazzale della Repubblica, 10, 65100 Pescara, Italy.}
\affiliation{$^4$ICRA -- International Center for Relativistic Astrophysics, Dipartimento di Fisica (G9), Universit\`a  di Roma, ``Sapienza'', Piazzale A. Moro, 5, 00185 Rome, Italy.}
\begin{abstract}
Quantization of systems with constraints can be carried on with several methods. In the Dirac formulation the classical generators of gauge transformations are required to annihilate physical quantum states to ensure their gauge invariance. Carrying on BRST symmetry it is possible to get a condition on physical states which, differently from the Dirac method, requires them to be invariant under the BRST transformation. Employing this method for the action of general relativity expressed in terms of the spin connection and tetrad fields with path integral methods, we construct the generator of BRST transformation associated with the underlying local Lorentz symmetry of the theory and write a physical state condition following from BRST invariance. This derivation grounds on the general results on the dependence of the effective action used in path integrals and consequently of Green's functions on the gauge fixing functionals used in the DeWitt - Faddeev - Popov method. \\
The condition we gain differs form the one obtained within Ashtekar's canonical formulation, showing how we recover the latter only by a suitable choice of the gauge fixing functionals. We finally discuss how it should be possible to obtain all the requested physical state conditions associated with all the underlying gauge symmetries of the classical theory using our approach. 
\end{abstract}
\maketitle
\section{Introduction}
The problem of quantization of constrained systems arises in many contexts of physical interest. The presence of constraints at a classical level avoids us to threat all the dynamical variables as independent ones, and entails several difficulties when we are to construct the quantum theory. In a program of canonical quantization which promotes all classical canonical variables to quantum operators one has to deal with the problem of imposing the constraints quantum mechanically. In the procedure \textit{\`{a} la Dirac} \cite{1} the constraint operators are imposed to annihilate physical states. This procedure stems from the observation that in the classical theory the constraint functions are generators of infinitesimal canonical transformations which don't alter the physical state of the system. \\
The Dirac procedure is widely used in different contexts, including quantization of general relativity \cite{2,3}. Nevertheless this procedure of quantization encounters several difficulties when we require the Dirac's conditions on physical states to be consistent with each other \cite{1,4} and the physical states selected by constraint operators to posses a finite scalar product allowing a probabilistic interpretation \cite{4,5}: moreover, in some cases this procedure can lead to a physical subspace of the entire Hilbert space that is curiously empty \cite{4}. Other difficulties arise when one tries to implement Dirac's procedure, which are not properly to ascribe to Dirac's theory for constrained systems, but to the canonical quantization framework this procedure is developed in. As a matter of fact, our experience on quantum field theory in special relativity showed us how canonical quantization methods, when applied to systems with infinite degrees of freedom, lead to several inconsistencies \cite{15,18}: for example, it is a remarkable fact that the Glashow - Weinberg - Salam theory for electroweak interactions cannot be consistently formulated by canonical quantization methods, while the only way it can be coherently written by is the Feynman's path integral. Even if Feynman's path integral can be derived after constructing the quantum theory by means of canonical quantization methods \cite{22}, such inconsistencies make necessary to postulate the path integral approach as a founding element of the quantum theory when we deal with systems with infinite degrees of freedom \cite{15}. 
It is for these reasons that we developed all of our work avoiding to use the Dirac procedure for constrained systems and canonical quantization methods at all, employing a method to derive conditions on physical states based on BRST symmetry and path integral methods uniquely. \vspace{2mm}\\
BRST symmetry \cite{4,6,7,8} was conceived at first within non-abelian gauge theories and showed to apply to a really wide class of systems of physical interest. Anyway in the literature there are different formulations for the BRST formalism, with substantial differences from each other. First of all, there exists a formulation of BRST symmetry for constrained systems based on canonical quantization methods which is widely diffused \cite{4,16}, being also employed in quantization of general relativity \cite{17}. Another approach \cite{6}, the one we followed in this work, to derive BRST symmetry is based entirely on path integral methods and is applicable to systems with infinite degrees of freedom avoiding those inconsistencies proper of canonical quantization methods we discussed above. The BRST method can then be extended to general gauge systems with open algebras employing a more sophisticated threatment \cite{24,26,25}.
We start with an enlightening and more or less known example, considering BRST symmetry for a non-abelian gauge theory. In order to compare path integral methods with canonical quantization ones, one can \cite{14} consider the N\"{o}ether charge following from BRST symmetry of the action and, taking an appropriate choice for the gauge fixing functionals in the DeWitt - Faddeev - Popov method, show it to be the generator of quantum BRST transformation within a canonical quantization framework. Otherwise, using solely path integral methods,  we show the BRST N\"{o}ether charge to generate quantum BRST transformation by means of Ward's identities, leaving the gauge fixing functionals completely arbitrary. Within this second approach the completely arbitrary gauge fixing functionals allow us to infer a physical state condition on states of the gauge fields following from BRST invariance.\\
Afterwards we turn our attention to general relativity expressed in first order formalism \cite{2,9}, in order to investigate the physicality condition for the states of the gravitational field arising from BRST invariance of the theory, following the same procedure employed for non-abelian gauge theories. In this procedure we will intentionally avoid to use canonical quantization methods. We are to determine a physical state condition on quantum states without thinking of classical hamiltonian constraints in order to compare, at the end of our calculation, our physicality condition required by BRST symmetry and derived with path integral methods with the one obtained using the Dirac quantization method employed within Ashtekar's canonical formulation \cite{2,5}. Comparing our physicality condition with the latter, we find they differ by an additional non-vanishing term. Finally we discuss the nature and possible reasons of this difference, showing how we recover the Dirac canonical condition in our BRST quantization only by a suitable choice of gauge fixing functionals within the DeWitt - Faddeev - Popov method. \vspace{2mm} \\
The paper is structured as follows. In Sec. \ref{sec:1} we discuss the BRST method in the general case, next in Sec. \ref{sec:2} we apply it to non-abelian gauge theories: in \ref{sec:9} we report the physical state condition obtained within the Dirac's procedure that we will compare in Section \ref{sec:8} with the one we gained with BRST symmetry. In Sec. \ref{sec:4} we apply this technique to general relativity and, after reporting in Section  \ref{sec:10} the physical states' condition we obtained with the Dirac's method, in Section \ref{sec:11} we discuss the physical state condition obtained for gravitational field's states with BRST symmetry. In Sec. \ref{sec:5} we discuss the differences between this physicality condition and the one's gained with the Dirac's procedure, showing how we can recover the latter only in a special case. 
\section{BRST symmetry in the general case}\label{sec:1}
Both general relativity expressed in first order formalism and non-abelian gauge theories are systems possessing an underlying symmetry under some infinitesimal transformations acting on fundamental fields $\phi^{r}$, which can be written as
\begin{equation}\label{eq:1}
\phi^{r} \rightarrow \phi^{r} + \epsilon^{A} \delta_{A} \phi^{r}\text{,}
\end{equation}
leaving the action $I\left[\phi \right]$ and integration measure
\begin{equation}\label{eq:4}
\left[ d \phi \right] \equiv \prod_{r} d \phi^{r}
\end{equation}
invariant. Following a generalization of the DeWitt - Faddeev - Popov method \cite{10,11} it is possible to show \cite{6} that the vacuum time-ordered product for generic operators $\mathcal{O}_{A}, \mathcal{O}_{B}, \cdots$ invariant under (\ref{eq:1}) can be written as 

\bea \label{eq:7}
\left\langle 0 \left| T\left( \mathcal{O}_{A} \mathcal{O}_{B} \cdots \right) \right| 0 \right\rangle & = &  \\ \nonumber
\frac{\int{\left[ d \phi \right] \left[ d h \right] \left[ d c^{\ast} \right] \left[ d c\right] \exp{\left(i I_{NEW}\left[ \phi, h, c, c^{\ast} \right]\right)}\mathcal{O}_{A}\mathcal{O}_{B} \cdots  \mathscr{B}\left[h\right]}}{\int{\left[ d \phi \right] \left[ d h \right] \left[ d c^{\ast} \right] \left[ d c\right] \exp{\left(i I_{NEW}\left[ \phi, h, c, c^{\ast} \right]\right)} \mathscr{B}\left[h\right]}}\text{,}
\eea
where the fields $h^{A}$ are known as `{Nakanishi - Lautrup}' fields \cite{12,13}, $c^{A}, c^{\ast A}$ are ghost and anti-ghost fields respectively, and $\mathscr{B}\left[ h \right]$ is the Fourier transform of some functionals $B\left[ f \right]$ we derive from the DeWitt - Faddeev - Popov theorem \cite{6}
\[
B\left[ f \right] = \int{\left[ d h \right] \exp{\left( i h^{A} f_{A} \right)} \mathscr{B}\left[h\right]} 
\]
and $I_{NEW}$ reads
\begin{equation}\label{eq:2}
I_{NEW}\left[ \phi, h, c, c^{\ast} \right] \equiv I\left[ \phi \right] + h^{A} f_{A} \left[ \phi \right] + c^{\ast B} c^{A} \delta_{A} f_{B} \left[ \phi \right] 
\end{equation}
for some arbitrary gauge fixing functionals $f_{A}\left[ \phi \right]$. The new total action $I_{NEW}$, depending on the functionals $f_{A}\left[ \phi \right]$, is not invariant under (\ref{eq:1}). In spite of this, it possesses a symmetry under an infinitesimal \textit{BRST transformation}, acting on a generic functional of the fields
$\psi_{i} = \left\{ \phi^{r}, h^{A}, c^{A}, c^{\ast A} \right\}$ as
\be \label{eq:5}
F\left[ \psi \right] \rightarrow F\left[ \psi \right] + \theta s F\left[ \psi \right]\text{,}
\ee
where $\theta$ is an ``{infinitesimal}'' Grassmann number and $s$ is the \textit{Slavnov operator}
\[
s \equiv c^{A} \delta_{A}\phi^{r} \frac{\delta_{L}}{\delta \phi^{r}} - \frac{1}{2}c^{B} c^{C}f_{\ \,BC}^{A} \frac{\delta_{L}}{\delta c^{A}}- h^{A} \frac{\delta_{L}}{\delta c^{\ast A}} \text{,}\]
where $\delta_{L}/{\delta{\psi_{i}}}$ denotes the left differentiation, defined by
\[
\delta F\left[ \psi \right] = \delta \psi_{i} \ \delta_{L} F\left[ \psi \right]/ {\delta \psi_{i}} \text{,}
\]
and the structure constants $f_{\ \, BC}^{A}$ are given by 
\be \label{eq:27}
\left[ \delta_{B},\delta_{C} \right] = f_{\ \, BC}^{A} \delta_{A} \text{.}
\ee 
Such a transformation acts on the Hilbert space through a \textit{BRST charge} $Q$ such that, given a generic functional $\Phi\left[ \psi \right]$, its variation under a BRST transformation is given by
\be \label{eq:12}
\delta_{\theta}\Phi\left[ \psi \right]  =  - i \left[ \theta Q , \Phi\left[ \psi \right] \right] \text{.}
\ee 
The condition on physical states given by the BRST symmetry can be obtained as follows. If we consider two physical states $\left| \alpha \right\rangle,\left| \beta \right\rangle$, the amplitude $\left\langle \left. \alpha \right| \beta \right\rangle$ can be expressed \cite{6} as a path integral evaluated with the action $I_{NEW}$. This amplitude must be independent \cite{27,28} on the gauge fixing functionals $f_{A}\left[ \phi \right]$ appearing in (\ref{eq:2}). This condition implies the invariance of physical states under BRST transformations
\begin{equation}\label{eq:3}
\left\langle \alpha \right| Q  = Q \left| \beta \right\rangle = 0
\end{equation}
which is the physicality condition we were searching for. \vspace{2mm}\\
We want to stress that in this Section we are dealing with gauge transformations with a closed algebra, as can be seen by (\ref{eq:27}): in some theories of physical interest the algebra is open, and closes only when the field equations are satisfied \cite{6}. In such theories it can be shown that the existence of some additional terms in (\ref{eq:27}) will make the operator $s^2$ not to vanish any more, requiring a more sophisticated treatment: the generalization of the formalism presented in this Section to theories with open gauge symmetry algebras was performed by Batalin and Vilkovisky \cite{24,25,26}. 
\section{Physical state condition for a non-abelian gauge theory}\label{sec:2}
\subsection{Phisicality condition according to Dirac's method}\label{sec:9}
It is well known that canonical quantization encounters some difficulties when applied to non-abelian gauge theories \cite{ 29,6,15,22} because of the existence of hamiltonian constraints. Anyway, employing the underlying gauge freedom of the theory, canonical quantization can be performed \cite{22}. Here we report the physical state condition for a non-abelian gauge theory with compact gauge group obtained within this canonical quantization framework employing the Dirac procedure, in order to compare it with the one we will obtain in Section \ref{sec:8}. 
An analysis of the classical action and implementation of Dirac - Bergman's algorithm leads us to the following primary
\[
\phi_{\alpha}(x) \equiv \Pi^{0}_{\phantom{0} \alpha}(x)
\] 
and secondary 
\be \label{eq:29}
\tilde{\phi}_{\alpha}(x) \equiv - \mathcal{D}_{a} F^{0a \alpha}(x)
\ee
first class constraints \cite{30,1}. If we choose the temporal gauge \cite{22} 
\be \label{eq:31}
A_{0}^{\phantom{0}\alpha}=0
\ee
we are left with the hamiltonian variables $\left\{A_{a}^{\phantom{a}\alpha}(x), \Pi^{a}_{\phantom{a} \alpha}(x)= -F^{0a\alpha}(x)\right\}_{a,\alpha,\vec{x}}$ satisfying the constraint (\ref{eq:29}) holding on the equations of motion. Following the Dirac procedure, if we call $n_{L}$ the gauge group's Lie algebra's dimension, the $2 n_{L} \infty^{3}$ physical degrees of freedom are obtained quantizing the canonical conjugated variables $\left\{A_{a}^{\phantom{a}\alpha}(x), \Pi^{a}_{\phantom{a} \alpha}(x)\right\}_{a,\alpha,\vec{x}}$ and imposing (\ref{eq:29}) 
\be \label{eq:30}
\mathcal{D}_{a} F^{0a \alpha}(x)\left| \psi \right\rangle = 0 \text{.}
\ee
We observe that (\ref{eq:30}) is not affected by operator ordering ambiguities because the only term containing the product of two operators is $f_{\gamma \beta}^{\phantom{\gamma \beta} \alpha} A_{a}^{\phantom{a} \gamma}F^{0a\beta}$
where, because of the structure constants' antisymmetry in all of their three indices ensured by gauge group's compactness \cite{32}, are present only products of $A_{a}^{\phantom{a} \gamma}$ and $F^{0a\beta}$ with $\gamma \neq \beta$ which, according to the equal time canonical commutation relations 
\[
\left[A_{a}^{\phantom{a} \alpha}(x), F^{0b\beta}(y) \right]= -i \delta^{(3)}\left(\vec{x}-\vec{y}\right)\delta_{a}^{b}\delta^{\alpha \beta} \text{,}
\] 
commute. 

\subsection{Phisicality condition according to BRST invariance}\label{sec:8}
We are now to apply the BRST method to a non-abelian gauge theory, in order to get a condition on physical states following from (\ref{eq:3}) to compare with (\ref{eq:30}). In this case we have $\left\{ \phi^{r}\right\}_{r} = \left\{ A_{\mu}^{\phantom{\mu} \alpha}(x) \right\}_{\mu,\alpha, x}$ and $I\left[ \phi \right]= S\left[ \mathcal{A} \right]$ where $S \left[\mathcal{A} \right]$ is the Yang-Mills action, and the transformations (\ref{eq:1}) are gauge transformations. Choosing the functional $\mathscr{B}\left[h \right]$ such that
\[
\mathscr{B}[h] = \exp{\left[ \frac{i\xi}{2 } \int{d^{4}x h_{\alpha}(x) h_{\alpha}(x)} \right]} \text{,}
\]
we are to determine the BRST generator $Q$.\\
If we set 
\[
I_{MOD}\left[ \psi \right]  \equiv  I_{NEW} \left[ \psi \right] + \frac{\xi}{2 } \int{d^{4}x h_{\alpha}(x)h_{\alpha}(x)} \text{,}
\]
using the explicit form of the transformation (\ref{eq:5}) it is possible to show that the integration measure $\mathcal{D} \psi $ is invariant under a generic \textit{local} BRST transformation where we take the Grassmann number $\theta$ to be a function of space-time coordinates $x$. Performing such a change of variable in the generating functional $Z\left[ j \right]$ evaluated with the ``{external currents}'' $j^{i}(x)$, we get
\begin{eqnarray*}
\int{\mathcal{D}\psi \exp{\left\{	i I_{MOD}\left[ \psi \right] + i \int{d^{4}x j^{i}(x) \psi_{i}(x)}\right\}}} &  = &  \\ \nonumber
\int{\mathcal{D}\psi \exp{\left\{i I_{MOD}\left[ \psi + \delta_{\theta(x)} \psi \right] + i \int{d^{4}x j^{i}(x) \left[ \psi_{i}(x) + \delta_{\theta(x)} \psi_{i}(x) \right]}\right\} }} +\\ 
+ O\left( \theta^{2} \right)\text{.}
\end{eqnarray*}
Retaining only the linear terms in $\theta(x)$, the latter equation enables us to derive the relation
\be \label{eq:6}
\partial_{\mu}^{x}\left\langle \mathscr{J}^{\mu}(x) \right\rangle_{j} + \left\langle s \psi_{i}(x) \right\rangle_{j}\mu^{i}(x) = 0\text{,}
\ee
where 
\be\label{eq:24}
\mu^{i}(x) \equiv \left\{ \begin{array}{ll}
j^{i}(x) & \psi_{i}(x) \ \text{bosonic}\\
- j^{i}(x) & \psi_{i}(x)  \ \text{fermionic}
\end{array}
\right. \text{,}
\ee
\be\label{eq:25}
\sigma^{i} \equiv \left\{ \begin{array}{ll}
1 & \psi_{i}(x) \ \text{bosonic}\\
- 1 & \psi_{i}(x)  \ \text{fermionic} 
\end{array}
\right.
\ee
and $\mathscr{J}^{\mu}$ is the N\"{o}ether current associated with the BRST symmetry for $I_{MOD}$
\[
\delta_{\theta(x)}I_{MOD}\left[ \phi, h, c, c^{\ast} \right]  = \int{d^{4}x \mathscr{J}^{\mu}(x)\partial_{\mu}^{x}  \theta  }\text{.}
\]
By means of (\ref{eq:6}) we will derive all of Ward's identities, which we will use to construct the BRST generator. If we calculate the functional derivatives of (\ref{eq:6}) with respect to the external currents $j^{i}$ and proceed by induction we get 
\begin{eqnarray}\label{eq:8}
0 & = & \partial_{\mu}^{x}\left\langle \psi_{i_{k}}\left( x_{k} \right) \cdots \psi_{i_{1}}\left( x_{1} \right) \mathscr{J}^{\mu}(x)  \right\rangle_{j=0}  - i \sum_{l=1}^{k} \sigma^{i_{1}} \cdots \sigma^{i_{l}} \langle \psi_{i_{k}}\left(x_{k} \right) \cdots  \\ \nonumber
& & \cdots \psi_{i_{l+1}}\left(x_{l+1}\right) s \psi_{i_{l}}(x)  \psi_{i_{l-1}}\left(x_{l-1}\right) \cdots \psi_{i_{1}}\left( x_{1} \right) \rangle_{j=0} \delta^{(4)}\left( x - x_{l} \right)  \text{.}
\end{eqnarray}
Using (\ref{eq:7}) we \textit{assume} the BRST current $\mathscr{J}^{\mu}$ to be conserved also quantum mechanically. This assumption does not imply any particular operator ordering for $\mathscr{J}^{\mu}$ because the only relation we are assuming true to proceed with our calculation is 
\be \label{eq:33}
\left\langle \psi_{i_{k}}\left( x_{k} \right) \cdots \psi_{i_{1}}\left( x_{1} \right) \partial_{\mu}^{x}\mathscr{J}^{\mu}  \right\rangle_{j=0} =0 \text{,}
\ee
where the BRST current appears exclusively as inserted into a Green's function. As far as this Green's function can be written as a path integral where the BRST current appear as a \textit{classical} quantity
\[
\left\langle \psi_{i_{k}}\left( x_{k} \right) \cdots \psi_{i_{1}}\left( x_{1} \right) \partial_{\mu}^{x}\mathscr{J}^{\mu}  \right\rangle_{j=0} = \int \mathcal{D} \psi \ e^{i I_{MOD}[\psi]} \psi_{i_{k}}\left( x_{k} \right) \cdots \psi_{i_{1}}\left( x_{1} \right) \partial_{\mu}^{x}\mathscr{J}^{\mu} \text{,}
\]
the assumption (\ref{eq:33}) is not affected by ordering ambiguities.\\
We can thus express (\ref{eq:8}) in terms of $k$-points Green's functions by means of the general rule \cite{19,20} to take time derivatives of time-ordered products, \emph{i.e.}
\begin{eqnarray}\label{eq:9}
0 & = & \sum_{l=1}^{k}\sigma^{i_{1}} \cdots \sigma^{i_{l}} \Big{\langle} 0 \Big{|} T \Big{\{} \psi_{i_{k}}\left(x_{k} \right)\cdots \psi_{i_{l+1}}\left(x_{l+1} \right) \Big{[} \left[\mathscr{J}^{0}(x), \psi_{i_{l}}\left(x_{l}\right) \right]_{\mp i_{l}} + \\ \nonumber
& &  - i \delta^{(3)}\left(\vec{x} - \vec{x}_{l} \right) s \psi_{i_{l}}(x) \Big{]} \psi_{i_{l-1}}\left(x_{l-1} \right) \cdots \psi_{i_{1}}\left(x_{1} \right) \Big{\}} \Big{|} 0 \Big{\rangle} \delta\left( x^{0} - x_{l}^{0} \right)  \text{.}
\end{eqnarray}
It is easy to see that if one supposes there is one time coordinate $x^{0}_{j}$ such that $x^{0}_{j} \neq x^{0}_{l} \ \forall  \ l \neq j, \ 1 \leq l \leq k$, from (\ref{eq:9}) follows 
\begin{eqnarray}
\label{eq:10}
0 & = & \Big{\langle} 0 \Big{|} T \Big{\{} \psi_{i_{k}}\left(x_{k} \right)\cdots \psi_{i_{j+1}}\left(x_{j+1} \right) \Big{[} \left[\mathscr{J}^{0}\left( x^{0}_{j}, \vec{x} \right), \psi_{i_{j}}\left(x_{j}\right) \right]_{\mp i_{j}} + \\ \nonumber
& &  - i \delta^{(3)}\left(\vec{x} - \vec{x}_{j} \right) s \psi_{i_{j}}\left( x^{0}_{j}, \vec{x} \right) \Big{]}  \psi_{i_{j-1}}\left(x_{j-1} \right) \cdots \psi_{i_{1}}\left(x_{1} \right) \Big{\}} \Big{|} 0 \Big{\rangle} \text{.}
\end{eqnarray}
Thus we see that the $k$-points Green's function in (\ref{eq:10}) a priori does not vanish identically, because it could be not  zero if we take $x^{0}_{j}$ equal to some other time coordinates $x^{0}_{l} \ \ l \neq j, \ 1 \leq l \leq k$, \emph{i.e.} it could have \textit{contact terms}. \textit{If} these contact terms are absent we can proceed and say that the Green's function given in (\ref{eq:10}) vanishes identically. Thus, as far as all the physical content of the theory is in the Green's functions and as far as the operator 
\[
\left[\mathscr{J}^{0}\left( x^{0}_{j}, \vec{x} \right), \psi_{i_{j}}\left(x_{j}\right) \right]_{\mp i_{j}} - i \delta^{(3)}\left(\vec{x} - \vec{x}_{j} \right) s \psi_{i_{j}}\left( x^{0}_{j}, \vec{x} \right)
\]
gives a vanishing contribution to a generic $k$-point Green function, we can proceed as if this operator vanishes
\be\label{eq:11}
\left[\mathscr{J}^{0}\left( x^{0}_{j}, \vec{x} \right), \psi_{i_{j}}\left(x_{j}\right) \right]_{\mp i_{j}} - i \delta^{(3)}\left(\vec{x} - \vec{x}_{j} \right) s \psi_{i_{j}}\left( x^{0}_{j}, \vec{x} \right)=0 \text{.}
\ee
Integrating (\ref{eq:11}) with respect to $\vec{x}$ we see that the charge $Q$ associated with the BRST current $\mathscr{J}^{\mu}$ is the generator of the BRST transformation on the Hilbert space
\[
\left[ \theta Q, \psi_{i}(x) \right] = i \delta_{\theta} \psi_{i}(x) \text{,}
\]
and thus it coincides with the charge $Q$ defined in (\ref{eq:12}). \\
Thus we have shown that the charge $Q$ associated with the BRST N\"{o}ether current $\mathscr{J}^{\mu}$ is the generator of the BRST transformation. We observe that this proof would lead to several difficulties if we'd use the canonical quantization formalism. In fact it is easy to see that in such a formalism more than one conjugate momenta to the fields $\psi_{i}$ would vanish identically, leading to difficulties in imposing the canonical commutation relations we need to  calculate the commutator in (\ref{eq:12}). 
\vspace{2mm}\\
We are now to derive a physical state condition following from the BRST invariance condition (\ref{eq:3}). Using the explicit form for the action $I_{MOD}$ and the definition of $\mathscr{J}^{\mu}$ as a N\"{o}ether current, it is easy to see that the BRST charge is given by 
\[
Q  =   \int d^{3}x \Big{[} - c^{\alpha}(x) \mathcal{D}_{a}F^{0 a \alpha}(x) + \text{\ terms depending on \ } f^{\alpha}\left( x; \phi \right)  \Big{]} \text{.}
\]
Looking at the physicality condition (\ref{eq:3}) and observing that the charge $Q$ is the sum of two terms, where the second one, differently from the first, depends \cite{27,28} on the completely arbitrary gauge fixing functionals $f_{A}\left[ \phi \right]$, we see that the first term must \textit{separately} annihilate physical states
\[
\int{d^{3}x c^{\alpha}(x) \mathcal{D}_{a}F^{0 a \alpha}(x)} \left| \psi \right\rangle = 0 \text{.}
\]
Observing that the ghost fields are necessarily all independent \cite{6} it is easy to see that the operator $ \mathcal{D}_{a}F^{0 a \alpha}(x)$ must annihilate physical states, i. e. 
\be \label{eq:13}
\mathcal{D}_{a}F^{0 a \alpha}(x)\left| \psi \right\rangle = 0  \text{.}
\ee
Thus \textit{the Gauss' constraint operator annihilates physical states}: this is the physicality condition we were searching for. This result is stated in the literature  \cite{23}, even if it's derived by a procedure different from the one followed here, and is well verified in perturbation theory \cite{15}, yielding the transverse polarization states of the particles associated with the gauge fields.  We emphasize that to obtain  (\ref{eq:13}) we never used any explicit expression for the constraints of the theory, since we avoided to employ the canonical quantization method nor the Dirac procedure, using exclusively the path integral and BRST quantization method. From this point of view we didn't impose the constraints to annihilate physical states as in the Dirac method, but \textit{derived} the Dirac condition for the Gauss' constraint, employing exclusively the BRST invariance of the theory. \vspace{2mm} \\
A priori (\ref{eq:13}) is affected by operator ordering ambiguities, since we're not giving a prescription on how to write the product of two field operators appearing in (\ref{eq:13}) in order to deal with the divergences of the fields. Therefore in the following will assume a given factor ordering in (\ref{eq:13}). Anyway this assumption does not affect the main result of our calculation: the constraint (\ref{eq:30}) differs from (\ref{eq:13}) because in Section \ref{sec:9} the field $A_{0}^{\phantom{0}\alpha}$ has been set to zero in the classical theory, so $F^{0 a \alpha}(x)$ in (\ref{eq:13}) has some additional terms with respect to $F^{0 a \alpha}(x)$ in (\ref{eq:30}) which are proportional to $A_{0}^{\phantom{0}\alpha}$. Employing the freedom to rearrange the products of operators in 
(\ref{eq:30}) explained in Section \ref{sec:9} we realize that the constraint (\ref{eq:30}) of the canonical theory reproduces exactly the first term in (\ref{eq:13}) not proportional to $A_{0}^{\phantom{0}\alpha}$, whatever the factor ordering we fixed in (\ref{eq:13}). Thus the constraint obtained employing BRST invariance differs from the one obtained with the Dirac procedure by some additional terms proportional to $A_{0}^{\phantom{0}\alpha}$. These terms, of course, will need for an appropriate ordering prescription, but their existence ensures that within the BRST method \textit{the canonical theory's Gauss' constraint does not annihilate physical states any more}. This is the main result of this Section which will find a deep analogy with the one gained for general relativity exposed in the following. 
\section{Physical state condition for general relativity}\label{sec:4}
\subsection{Phisicality condition according to Dirac's method}\label{sec:10}
Following the same procedure showed in Section \ref{sec:8} for a non-abelian gauge theory, by a suitable gague choice the gravitational field's action can be cast in canonical form and Poisson brackets turned into canonical commutators. As a matter of fact, employing the gauge freedom to redefine locally the tetrad fields $e_{I\mu}$ \cite{2,5,9}, in this Subsection we will choose the \textit{temporal gauge} \cite{31} 
\be \label{eq:34}
e_{0}^{\mu}=n^{\mu} \text{,}
\ee
analogous to (\ref{eq:31}). With this choice the gravitational field's action can be cast into a Legendre's transform \cite{2,5}, and the canonical conjugated variables $\left\{ A_{i}^{a}(x), E^{a}_{i}(x) \right\}_{a,i,\vec{x}}$ satisfy the canonical commutation relations 
\[
\left[ A_{i}^{a}(x), E^{b}_{j}(y) \right] = i \delta_{b}^{a} \delta_{j}^{i} \delta^{(3)}(\vec{x}-\vec{y}) \text{.}
\]
Imposing the classical secondary constraint $\mathcal{D}_{a}  E^{a}_{i}(x)=0$ by the Dirac procedure, we find
\be \label{eq:32}
\mathcal{D}_{a}  E^{a}_{i}(x) \left| \psi \right\rangle =0 \text{,}
\ee
which is the analogous of (\ref{eq:30}). Following the same argument of Section \ref{sec:8} we see that, because of the antisymmetry of $SU(2)$'s structure constants, (\ref{eq:32}) is not affected by operator ordering ambiguities. 

\subsection{Phisicality condition according to BRST invariance}\label{sec:11}
In order to employ BRST symmetry for general relativity we consider the Einstein - Hilbert action expressed in first order formalism \cite{2,9}. Here we point we used the tetrad variables instead of the metric tensor because the first order formalism which stems from the tetradic formulation makes clear the analogy between general relativity and gauge theories, with the proper Lorentz group identified with the gauge group. By means of this identification we developed all of this Section's calculations in analogy with those we made in Section \ref{sec:2} for a non-abelian gauge theory. The lagrangian dynamical variables are the tetrad field $e^{I}_{\mu}$ and the spin connection $\omega_{\mu J}^{I}$ and the action reads \cite{9,31}
\[
S_{-}[e,\omega] =  \frac{i}{16 \pi G} \int{d^{4}x \epsilon^{\mu \nu \rho \sigma} e_{I \rho} e_{J \sigma}R[\omega_{-}]_{\mu \nu}^{IJ}}\text{.}
\]
Setting 
\[
\left\{ \phi_{r} \right\}_{r} \equiv \left\{ \left\{ e^{I}_{\mu}(x) \right\}_{I,\mu,x}, \left\{ \omega_{\alpha \mu}(x) \right\}_{\alpha, \mu, x} \right\}
\]
it is easy to see that the action and measure are invariant under the infinitesimal proper local Lorentz transformations
\begin{eqnarray*}
\left\{\begin{array}{lll} e^{I}_{\mu}(x) & \rightarrow & \Lambda^{I}_{\phantom{I} J}(x)e^{J}_{\mu}(x) \\
\omega_{\mu J}^{I}(x) &  \rightarrow & \Lambda^{I}_{\phantom{I} K}(x)\omega_{\mu L}^{K}(x)\left[ \Lambda(x)^{-1}  \right]^{L}_{\phantom{L} J} - \left[ \partial_{\mu} \Lambda^{I}_{\phantom{I} K}(x)  \right]   \left[ \Lambda(x)^{-1}  \right]^{K}_{\phantom{K} J} 
\end{array}\right.\text{,}
\end{eqnarray*}
 which can be written in the form  (\ref{eq:1}). The N\"{o}ether current $\mathscr{J}^{\mu}$ associated with the BRST symmetry for $I_{NEW}$ is now given by 
\be \label{eq:14}
\delta_{\theta(x)}I_{NEW}\left[ \phi , h, c, c^{\ast} \right] = \int{d^{4}x \sqrt{-g(x)} \mathscr{J}^{\mu}(x) \partial_{\mu} \theta(x) } 
\ee
and the BRST charge reads
\be\label{eq:17}
Q \equiv \int{d^{3}x \sqrt{-g(x)} \mathscr{J}^{0}(x) } \text{.}
\ee
It is easy to show that, using (\ref{eq:14}) and the expression for the infinitesimal variation of $\omega_{\mu J}^{I}$ under local gauge transformations, the BRST charge can be expressed in terms of the Ashtekar connection $A^{i}_{a}$ according to
\bea\label{eq:18}
Q & = & \int d^{3}x\Bigg{\{} - \frac{i}{8 \pi G} \left[ \epsilon_{jk}^{\phantom{jk} l} c_{\underset{j}{-}}(x)   A^{k}_{a}(x) - \partial_{a}^{x} c_{\underset{l}{-}}(x) \right] e_{I b}(x) e_{J c}(x) \epsilon^{a b c}  \times \\ \nonumber
 & & \times T_{\ l}^{- IJ}  +  \text{\ terms depending on \ }   f^{\alpha}\left( x ; \phi \right)\Bigg{\}}
\eea
where $T^{-IJ}_{\ k}$ are the generators of the self-dual part of $\mathfrak{so}(3,1; \mathbb{C} )$. As we expected on general grounds \cite{27,28}, the BRST charge depends on the gauge fixing functionals used in the DeWitt - Faddeev - Popov method. 
\vspace{2mm}\\
We are now to show that the N\"{o}ether charge $Q$ in the quantum theory coincides with the BRST generator defined in (\ref{eq:12}). To do this we will again make use of path integral methods and employ Ward's identities. Setting 
\[
\left\{ \psi_{i}(x) \right\}_{i} \equiv \left\{ \left\{ e^{I}_{\mu}(x) \right\}_{I, \mu},    \left\{ \omega_{\alpha \mu}(x) \right\}_{\alpha, \mu},\left\{ h_{\alpha}(x) \right\}_{\alpha}, \left\{ c_{\alpha}(x) \right\}_{\alpha},   \left\{ c^{\ast}_{\alpha}(x) \right\}_{\alpha}   \right\}\text{,}
\]
and using (\ref{eq:24}), (\ref{eq:25}) we can show the integration measure $\mathcal{D}\psi$  to be still invariant under infinitesimal local BRST transformations, so that we get the following identity holding for the generating functional
\begin{eqnarray}\label{eq:23}
\int{\mathcal{D}\psi \exp{ \left\{	i I_{NEW}\left[ \psi \right] + i \int{d^{4}x j^{i}(x) \psi_{i}(x)}\right\}}}& = & \\ \nonumber
 \int\mathcal{D}\psi \exp \bigg{\{} i I_{NEW}\left[ \psi + \delta_{\theta(x)} \psi \right] + i \int d^{4}x j^{i}(x) \big{[} \psi_{i}(x) + \\ \nonumber 
 + \delta_{\theta(x)} \psi_{i}(x) \big{]} \bigg{\}}+  O\left( \theta^{2} \right)\text{.}
\end{eqnarray}
Here we want to comment on the meaning and definition of the integration measure $\mathcal{D}\psi \equiv \prod_{i,x} d\psi_{i}(x)$ given by (\ref{eq:4}): in this case $\mathcal{D}\psi$ is defined just as in Section \ref{sec:2}, which is the usual way to define the field integration measure in special relativistic quantum field theory by a suitable partition of the coordinate domain $\mathbb{R}^{4}$ into a discrete set of points. This definition has been also employed for gravitational field's path integrals in previous works \cite{21} and, even though borrowed from a special relativistic context, can be carried on even with a non vanishing gravitational field, since it employs only the existence of a given coordinate system $\left\{ x^{\mu} \right\}_{\mu}$ on the spacetime manifold $\mathcal{M}$, which can be always introduced. In particular, the statement that the measure $\mathcal{D}\psi$ is ill-defined if we are dealing with a non vanishing gravitational field because in this case the intervals $ds^{2}$, and consequently the spacetime's lattice structure, depends on the configuration of the metric $g_{\mu \nu}$ is not correct, since the definition we gave for $\mathcal{D}\psi$ employs a spacetime lattice structure independent on the metric configuration and given in terms a suitable partition the coordinates' domain only. \\
From (\ref{eq:23}) we get the following relation generating all of Ward's identities 
\[
\partial_{\mu} \left\langle \sqrt{-g(x)}\mathscr{J}^{\mu}(x)  \right\rangle_{j}  + \left\langle  s\psi_{i}(x)  \right\rangle_{j} \mu^{i}(x) = 0  \text{.}
\]
According to (\ref{eq:1}), to reproduce $k$-points Green's functions we have to consider a path integral involving the weighting functionals $\mathscr{B}[h]$. Proceeding by induction it is possible to show that the following relation holds $\forall k \geq0$
\begin{eqnarray*}
\partial_{\mu}^{x} \left\langle \mathscr{B}\left[ h \right]  \psi_{i_{k}}\left( x_{k} \right) \cdots \psi_{i_{1}}\left( x_{1} \right) \sqrt{-g(x)} \mathscr{J}^{\mu}(x) \right\rangle_{j=0} + \\  \nonumber
- i \sum_{l=1}^{k}\sigma^{i_{1}} \cdots \sigma^{i_{l}} \big{\langle} \mathscr{B}\left[ h \right] \psi_{i_{k}}\left( x_{k} \right) \cdots \psi_{i_{l+1}} \left( x_{l+1} \right) s \psi_{i_{l}}(x) \psi_{i_{l-1}} \left( x_{l-1} \right) \cdots \\ \nonumber
\cdots \psi_{i_{1}}\left( x_{1}\right) \big{\rangle}_{j=0}
\delta^{(4)}\left( x - x_{l} \right) &  = &  0 \text{.}
\end{eqnarray*}
Following the same arguments of Section \ref{sec:1} and choosing
\[
\mathscr{B}\left[ h \right] = \exp{\left(  \frac{i}{2 \xi}  h^{A} h^{A} \right)}\text{,}
\]
we obtain the following identity for the Green's functions
\begin{eqnarray}\label{eq:15}
\sum_{l=1}^{k}\sigma^{i_{1}} \cdots \sigma^{i_{l}}  \Bigg{\langle} 0 \Bigg{|} T\Bigg{\{} \psi_{i_{k}}\left( x_{k} \right) \cdots \psi_{i_{l+1}}\left( x_{l+1} \right) \times \\ \nonumber
\times \Bigg{[} \left.\left[ \sqrt{-g(x)} \mathscr{J}^{0}(x) , \psi_{i_{l}}\left( x_{l}  \right)  \right]_{\mp i_{l}}\right|_{x^{0}= x^{0}_{l}} - i \delta^{(3)}\left( \vec{x} - \vec{x}_{l} \right) s \psi_{i_{l}}(x)  \Bigg{]} \times \\ \nonumber
\times \psi_{i_{l-1}}\left( x_{l-1} \right) \cdots \psi_{i_{1}}\left( x_{1} \right)  \Bigg{\}} \Bigg{|} 0 \Bigg{\rangle} \delta\left( x^{0} - x^{0}_{l} \right)&  =&  0  \text{.}
\end{eqnarray}
To avoid confusion, here we comment on the meaning and definition of the $T$ product in (\ref{eq:15}), following an argument similar to the one we used above to comment on the field integration measure's definition. The temporal ordering used in (\ref{eq:15}) and in the following is the straight generalization of the temporal ordering used in special relativistic quantum field theory to define Green's functions: the operators $\mathcal{O}(x_{1}), \cdots , \mathcal{O}(x_{n})$ are ordered by means of the time coordinate $x^{0}$ referred to a given coordinate system on the spacetime manifold $\mathcal{M}$. This definition can be carried on also in general relativity, since it refers only to a given coordinate system $\left\{ x^{\mu} \right\}_{\mu}$ on $\mathcal{M}$, which is always possible to introduce. In particular, the statement that the $T$ product given in (\ref{eq:9}) is ill-defined when we are dealing with a non vanishing gravitational field because in this case the time intervals, depending on the metric $g_{\mu \nu}$, are not uniquely given is not correct, because the $T$ product we are dealing with in not referred to the metric configuration in any way and employs spacetime time coordinates only. \\
Proceeding as in Section \ref{sec:2}, \textit{if} contact terms are absent \textit{and} the BRST current is conserved in the quantum theory, following the same arguments given in Section (\ref{sec:8}) to justify the absence of operator ordering assumptions following from BRST current's conservation, we can use the general rule for the time derivative of a time ordered product expressed in terms of equal-times (anti)commutators and say that the Green's function in (\ref{eq:15}) vanishes identically
\begin{eqnarray}\label{eq:16}
\Bigg{\langle} 0 \Bigg{|} T\Bigg{\{} \psi_{i_{k}}\left( x_{k} \right) \cdots \psi_{i_{j+1}}\left( x_{j+1} \right) \Bigg{[} \left.\left[ \sqrt{-g(x)} \mathscr{J}^{0}(x) , \psi_{i_{j}}\left( x_{j}  \right)  \right]_{\mp i_{j}}\right|_{x^{0}= x^{0}_{j}} + \\ \nonumber
 - i \delta^{(3)}\left( \vec{x} - \vec{x}_{j} \right) s \psi_{i_{j}}(x^{0}_{j},\vec{x}) \Bigg{]}  \psi_{i_{j-1}}\left( x_{j-1} \right) \cdots \psi_{i_{1}}\left( x_{1} \right)  \Bigg{\}} \Bigg{|} 0 \Bigg{\rangle}&  = & 0  \text{.}
\end{eqnarray}
As far as all the physical content of the theory is in the Green's functions, we arrive at the conclusion that the operator in square brackets in (\ref{eq:16}) vanishes identically, which shows that the N\"{o}ether charge $Q$ defined in (\ref{eq:17}) generates the BRST transformation in the Hilbert space.
\vspace{2mm}\\
Using the expression (\ref{eq:18}) and the physicality condition (\ref{eq:3}) we see that, as far as the gauge fixing functionals $f^{\alpha}$ are completely arbitrary, the condition (\ref{eq:3}) can be satisfied only if \be \label{eq:19}
\int d^{3}x{ c_{\underset{j}{-}}(x) \mathcal{D}_{a} \left[ e_{Ib}(x) e_{Jc}(x) \epsilon^{ a b c } T^{- IJ}_{\ j}  \right]} \left| \psi \right\rangle = 0  
\ee
where $\mathcal{D}_{a}$ is the  $SU(2)$ covariant derivative. Being the ghost fields all independent, (\ref{eq:19}) leads to 
\be \label{eq:20}
\mathcal{D}_{a} \left[ e_{Ib}(x) e_{Jc}(x) \epsilon^{ a b c } T^{- IJ}_{\ j}  \right] \left| \psi \right\rangle = 0
\ee
which is the physicality condition we were searching for. To compare (\ref{eq:20}) with the usual physicality condition given by Gauss' constraint used in Ashtekar's canonical formulation, we use the explicit expression for the generators $T^{-IJ}_{\ k}$ and cast (\ref{eq:20}) into the following final form 
\begin{equation}\label{eq:21}
\mathcal{D}_{a} \left[ E^{a}_{j}(x) + i e_{j b}(x) e_{0 c}(x) \epsilon^{a b c}  \right] \left| \psi \right\rangle = 0  \text{.}
\end{equation}
As observed for (\ref{eq:30}), (\ref{eq:21}) is affected by operator ordering ambiguities, and we are assuming a given  ordering. Anyway, as observed in Section \ref{sec:9}, this factor ordering problem does not affect our main statement. As a matter of fact, employing the freedom to rearrange operator products in (\ref{eq:32}) explained in Section \ref{sec:10}, we see that the constraint (\ref{eq:32}) derived within Dirac's procedure coincides exactly with the first addend in (\ref{eq:21}). Thus, even if we are not givin any prescription to manage the ordering of the additional term $i \mathcal{D}_{a}\left[  e_{j b}(x) e_{0 c}(x) \epsilon^{a b c}\right]$ in (\ref{eq:21}), we showed that \textit{within the BRST method the usual Gauss' constraint obtained with the Dirac's procedure does not annihilate physical states any more}.  This must be considered the main result of this paper. Reasons and consequences of this difference with the Dirac's formulation will be discussed in the following. 
\section{Discussion and conclusions}\label{sec:5}
We employed BRST symmetry for general relativity expressed in first order formalism and gained, using path integral methods and BRST invariance, a physical state condition for the gravitational field's states which avoids the difficulties and inconsistencies raising in the Dirac procedure in imposing the classical constraints in the quantum theory. We want to stress how our derivation of such physical states' condition, both in Section \ref{sec:2} and \ref{sec:4}, employs the dependence of the action $I_{MOD}$ used in path integrals and of Green's functions on the gauge fixing functionals, illustrated for the first time in \cite{27} and developed within the Batalin - Vilkovisky method in \cite{28}: it is by means of this dependence that we found that the  BRST generator and the BRST invariance condition (\ref{eq:3}) to contain the gauge fixing functionals. Thus, once the gauge dependent part of $Q$ in (\ref{eq:3}) was eliminated, we were in the position to derive the final results (\ref{eq:13}) and (\ref{eq:21}). The condition (\ref{eq:21}) we found differs from the usual of Ashtekar's canonical formulation by an additional term. This additional term contains the operator $e_{0a}$, which is usually set to zero in the classical theory in Ashtekar's 
canonical formulation by the gauge condition (\ref{eq:34}). In our formulation the operator $e_{0a}$ in general cannot be set to zero, because it's treated as a lagrangian dynamical variable of the theory, on the same footing of the other components of the tetrad. Thus the additional term we found in (\ref{eq:21}), even if affected by operator ordering ambiguities is, in general, not vanishing, nor can be set to zero by a suitable gauge fixing as could be done in the classical theory. This raises the question of the equivalence of two gauge fixing procedures  performed before and after quantization: we proved that, as far the physical state condition is concerned, they lead to substantially different results. We want to stress we choose a quantization procedure which avoids to make any kind of explicit gauge fixing procedure: a gauge fixing done before quantization can break, at least formally, the underlying gauge symmetry of the theory, and in general it is not obvious if this symmetry is preserved after the quantization procedure. For example in the quantum theory of the electromagnetic field it can be particularly useful to perform quantization in Coulomb or temporal gauge, although the transverse condition $\vec{\nabla} \cdot \vec{A}=0$ or (\ref{eq:31}) are not Lorentz-invariant. In our case, a gauge fixing procedure like the one's performed in Ashtekar's canonical formulation breaks a fundamental physical symmetry of four-dimensional space, such as the local Lorentz symmetry.
\vspace{2mm}\\
Even if our quantization procedure can seem to be inequivalent to Ashtekar's one we observe that, by a suitable choice of the gauge fixing functionals, there exists a formal limit in which Ashtekar's condition can be reproduced. If we choose the gauge fixing functionals such that $f^{\alpha}(\phi;x)=0$ implies the temporal gauge condition, and take the limit $\xi\rightarrow \infty$, integrating over the fields $h_{\alpha}(x)$ in a generic Green's function we obtain an integrand factor of the form 
\be \label{eq:22}
\exp \left[-\frac{i \xi}{2} \int{d^{4}x f^{\alpha}(x;\phi)  f^{\alpha}(x;\phi)   }   \right]\text{.}
\ee 
According to this weighting factor, being $\xi \rightarrow \infty$, we see that the unique regions in the space of fields' configuration that give a non vanishing contribution to a generic Green's function are those where $f^{\alpha}(x;\phi)=0 $, \emph{i.e.} those where the temporal gauge condition is satisfied. Thus, if we consider a Green's function containing the operator $e_{0a}$, the only regions in fields' configuration space that give a non vanishing contribution to such a Green's function are those where $e_{0a}(x) =0$. Thus this Green's function vanishes identically in this limit. As far as all the content of the theory is in Green's functions, in this limit we can take the operator $e_{0a}$ to vanish and, according to (\ref{eq:21}), recover the physical states' condition of Ashtekar's canonical's formulation. The reason for which we are forced to take $\xi \rightarrow \infty$ to reproduce this condition is that in the quantum theory \text{all} of the fields' configurations contribute to a generic Green's function when we integrate over them to calculate vacuum expectation values of time-ordered products. Thus in the quantum theory it is not sufficient to take $f^{\alpha}$ reproducing the temporal gauge condition to ensure that $e_{0a}$ vanish identically, because all of the fields' configurations, included those where $e_{0a}\neq0$ give contributions, resulting in a non vanishing Green's functions containing $e_{0a}$. Anyway, taking $\xi \rightarrow \infty$ we select only those regions where  $e_{0a}=0$ and force $e_{0a}$ to vanish identically. \\
We observe that such a limiting procedure, even if the physical amplitudes $\left\langle \alpha \left| \right. \beta \right\rangle$ do not depend on the gauge fixing functionals nor on $\xi$, cannot be said to be equivalent to any other configuration for the gauge fixing functionals in such a way that one could definitively choose the physical state condition given by Gauss' constraint to hold anyway. In fact this limiting procedure could be reached only asymptotically, and be mathematically ill-defined. This can be clearly seen observing that this limiting procedure implies that a generic Green's function containing $e_{0a}$ vanishes identically only if we assume that the limiting operation can be exchanged with the integration over fields' configuration one, so that one can state that the phase in (\ref{eq:22}) oscillates rapidly, excepted when $f^{\alpha}(\phi;x)=0$. This assumption can clearly be proved to be valid only under some suitable regularity condition, and in general it turns out to be a not trivial result.  \vspace{2mm} \\
We conclude observing that a future perspective for this work is to reproduce \textit{all} the physical state conditions following by each of the underlying gauge symmetries of the theory employing  this BRST method. In particular, using as gauge symmetry (\ref{eq:1}) the general coordinate transformation symmetry we expect to get a physical state condition corresponding to the diffeomorphism constraints \cite{2,5,9,31}
\be \label{eq:26}
\mathcal{H}  =  \frac{i}{2 ^{(3)}e} \epsilon_{i}^{\phantom{i}jk} E_{j}^{b} E_{k}^{c} F[A]_{bc}^{i} \text{,} \ \mathcal{H}_{a}  = -  E_{i}^{b} F[A]_{a b}^{i}\text{.}
\ee
In other words, as we got the physical state condition (\ref{eq:20}) implementing Lorentz invariance, if we employ diffeomorphism invariance of the theory we expect to gain some constraints on physical states that would be the analogous of (\ref{eq:26}) in Ashtekar's theory like so (\ref{eq:20}) is the analogous of Ashtekar's Gauss' constraint 
(\ref{eq:32}). Anyway, this procedure would require a much more sophisticated treatment, stemming from the fact that in this case the integration measure (\ref{eq:4}) is not invariant under (\ref{eq:1}) any more, and it may become necessary to have recourse to a non trivial measure definition which has revealed unnecessary for our present treatment, where we are concerned with the local Lorentz symmetry alone. 
\section*{Acknowledgments}
We would like to thank M. Testa for his precious suggestions on the development of this work: it was him to suggest us to use Ward's identities to show the BRST N\"{o}ether charge to coincide with the BRST generator, which revealed to be a fundamental step to develop all of our work. Then we would like to thank L. Maiani for his skillfulness in commenting and giving suggestions on our use of path integral methods and Ward's identities.  Finally, we would like to thank O.M. Lecian too for her advices on some technical steps of our scheme.

\end{document}